\documentclass[
    ,final            
  ]
  {aipproc}

\layoutstyle{6x9}

\begin{document}

\title{On the role of injection in kinetic approaches \\
to nonlinear particle acceleration \\
at non relativistic shocks}


\author{S. Gabici}{
  address={Max--Plank--Institut f\"ur Kernphysik, Saupfercheckweg 1, 69117 Heidelberg, Germany}
}

\author{P. Blasi}{
  address={INAF/Osservatorio Astrofisico di Arcetri, Largo E. Fermi 5, 50125 Firenze, Italy}
}

\author{G. Vannoni}{
  address={Max--Plank--Institut f\"ur Kernphysik, Saupfercheckweg 1, 69117 Heidelberg, Germany}
}

\begin{abstract}
The dynamical reaction of the particles accelerated at a shock front
by the first order Fermi process can be determined within kinetic models 
that account for both the hydrodynamics of the shocked fluid and the transport 
of the accelerated particles. These models predict the appearance of multiple 
solutions, all physically allowed. We discuss here the role of injection in 
selecting the {\it real} solution, in the framework of a simple phenomenological 
recipe, which is a variation of what is sometimes referred to as 
{\it thermal leakage}. 
In this context we show that multiple solutions basically disappear and when
they are present they are limited to rather peculiar values of the parameters.  
\end{abstract}

\maketitle

Diffusive shock acceleration is thought to be responsible for acceleration of cosmic rays in several astrophysical environments. Despite the success of this theory, some issues are still subjects
of much debate, for the theoretical and phenomenological implications that 
they may have. One of the most important of these is the reaction of the 
accelerated particles on the shock: the violation of the 
{\it test particle approximation} occurs when the acceleration process 
becomes sufficiently efficient that the pressure of the accelerated
particles is comparable with the incoming gas kinetic pressure.
Both the spectrum of the particles and the structure of the shock
are changed by this phenomenon, which is therefore intrinsically 
nonlinear (Ellison, these proceedings).
Nonlinear effects in shock acceleration of thermal particles result in 
the appearance of multiple solutions in certain regions of the parameter
space. This phenomenon is very general and was found in both the two-fluid
 \cite{duefluidi} and kinetic models 
\cite{malkov,pasquale2,noi}. 
Here we investigate the phenomenon of multiple solutions and show that the appearance of these solutions is dramatically reduced if a self consistent model for injection is adopted. 

\section{A semi--analytical approach to the problem}

Following the approach presented in \cite{pasquale1,pasquale2,noi}, we solve the steady-state transport equation for the cosmic ray distribution function $f(x,p)$ at a plane shock wave: 
\begin{equation}
\frac{\partial}{\partial x}
\left[ D  \frac{\partial}{\partial x} f(x,p) \right] - 
u  \frac{\partial f (x,p)}{\partial x} + 
\frac{1}{3} \frac{d u}{d x}~p~\frac{\partial f(x,p)}{\partial p} + Q_0(p) \delta(x) = 0
\label{eq:trans}
\end{equation}
coupled with the continuity and Euler equations describing the dynamics of the flow:
\begin{equation}
\rho_0 u_0 = \rho u ~~~ ; ~~~ \rho_0 u_0^2 + P_{g,0} = \rho u^2 + P_{g} + P_{CR}.
\label{eq:fluid}
\end{equation}
All the symbols have their usual meanings, and the subscript 0 refers to quantities measured upstream infinity. $P_{g}$ and $P_{CR}$ represent the contributions to the total pressure of thermal gas and cosmic rays.
We introduce the quantity $u_p$ defined as:
\begin{equation}
u_p = u_1 - \frac{1}{f_0} \int_{-\infty}^{0^-} dx \frac{d u}{d x} f(x,p),
\label{eq:up}
\end{equation}
whose physical meaning is instrumental to understand the nonlinear 
reaction of particles. The function $u_p$ is the average fluid 
velocity experienced by particles with momentum $p$ while diffusing 
upstream away from the shock surface. In other words, the effect of the 
average is that, instead of a constant speed $u_1$ upstream, a particle 
with momentum $p$ experiences a spatially variable speed, due to the 
pressure of the accelerated particles. Since 
the diffusion coefficient is in general $p$-dependent, particles with 
different energies {\it feel} a different compression coefficient, higher 
at higher energies if, as expected, the diffusion coefficient is an 
increasing function of momentum.

It can be shown that equations \ref{eq:trans} and \ref{eq:fluid} can be reduced to an integral-differential equation for the quantity $U(p) = u_p/u_0$, that when solved with the boundary condition $U(p_{max})=1$ \footnote{This corresponds to assuming that the fluid is not affected by the cosmic ray pressure at upstream infinity.} provides us with the functions $u_p$ and $f_0(p)$ describing the flow profile and the particle distribution function at the shock \cite{pasquale1,pasquale2,noi}.

\begin{figure}
\label{fig:sol}
\includegraphics[height=.3\textheight]{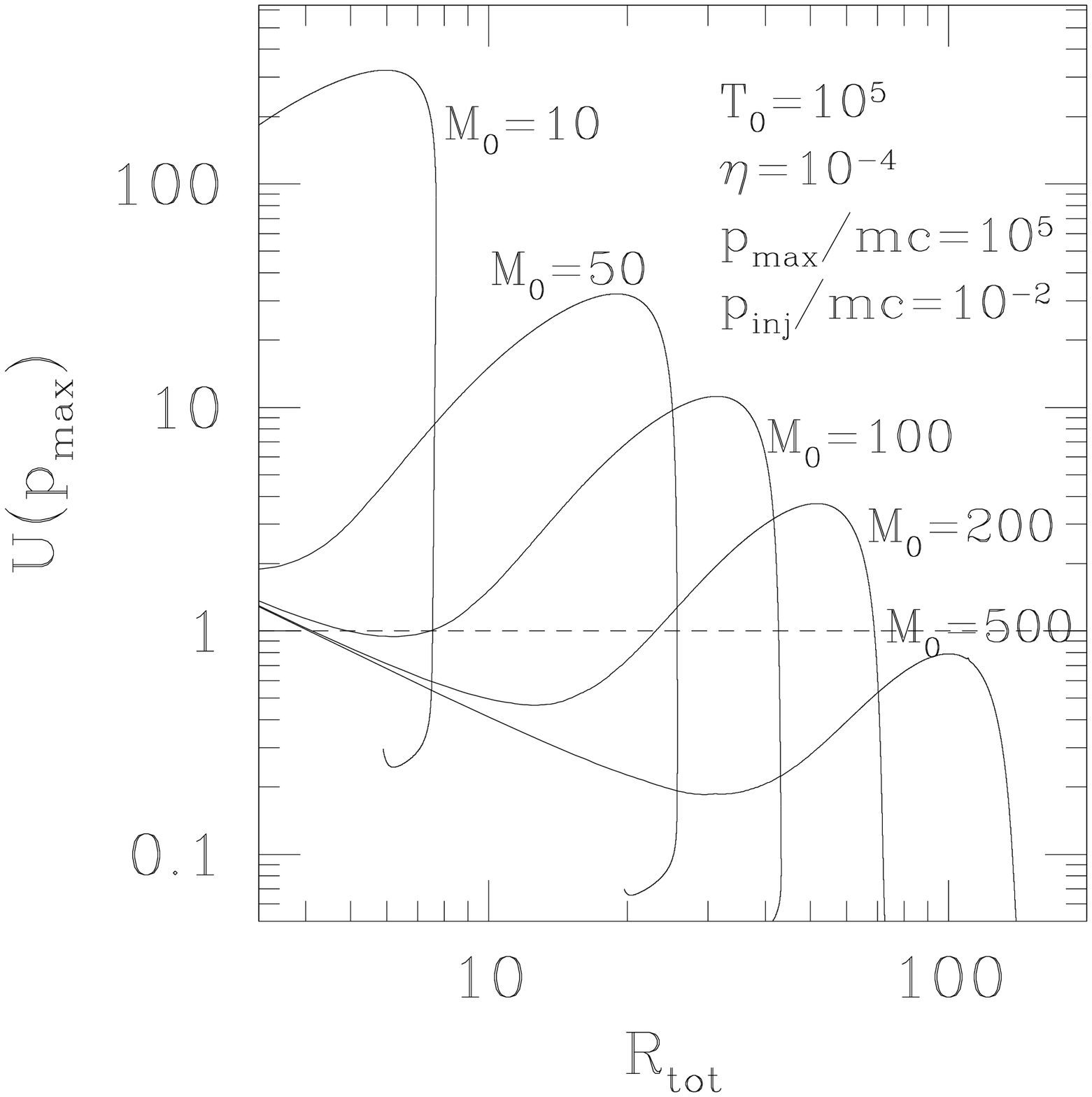}
\includegraphics[height=.3\textheight]{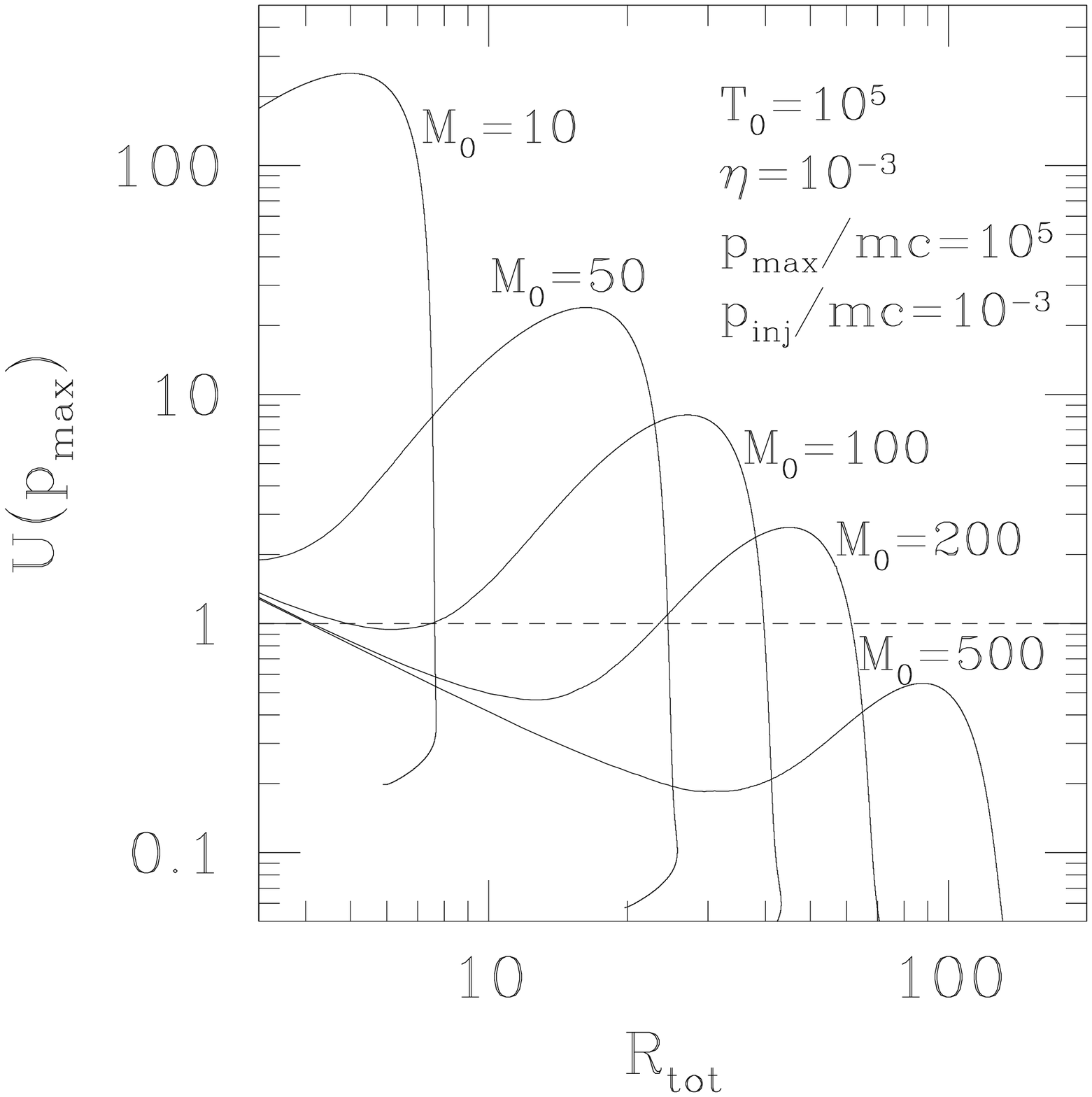}
\caption{$U(p_{max})$ as a function of the total compression 
factor} 
\end{figure}

In the problem described above there are several 
independent parameters. While the Mach number of the shock and the 
maximum momentum of the particles are fixed by the physical conditions
in the environment, the injection momentum $p_{inj}$ and the acceleration efficiency $\eta$
are free parameters.
The procedure to be followed to determine the solution 
was defined in \cite{pasquale2,noi}.
In Fig. 2 
we plot $U(p_{max})$ as a function of the total compression factor of the shock $R_{tot}$, 
for $T_0=10^5 K$, $p_{max}=10^5 mc$ and $p_{inj}=10^{-2} mc$ 
in the left panel and $p_{inj}=10^{-3} mc$ in the right panel 
($m$ here is the mass of protons). The parameter 
$\eta$ is $10^{-4}$ in the left panel and $10^{-3}$ in the right panel.
The different curves refer to different choices of the Mach number
at upstream infinity. The physical solutions are those corresponding 
to the intersection points with the horizontal line $U(p_{max})=1$, so that 
multiple solutions occur for those values of the parameters for 
which there is more than one intersection with $U(p_{max})=1$. These 
solutions are all physically acceptable, as far as the conservation 
of mass, momentum and energy are concerned. Fig.\ref{fig:sol} shows that multiple solutions are a very common phenomena, since they are found if very different values of the parameters are adopted. In the next section we show how the situation changes if a better description of injection is implemented in the model.

\begin{figure}
\includegraphics[height=.65\textheight]{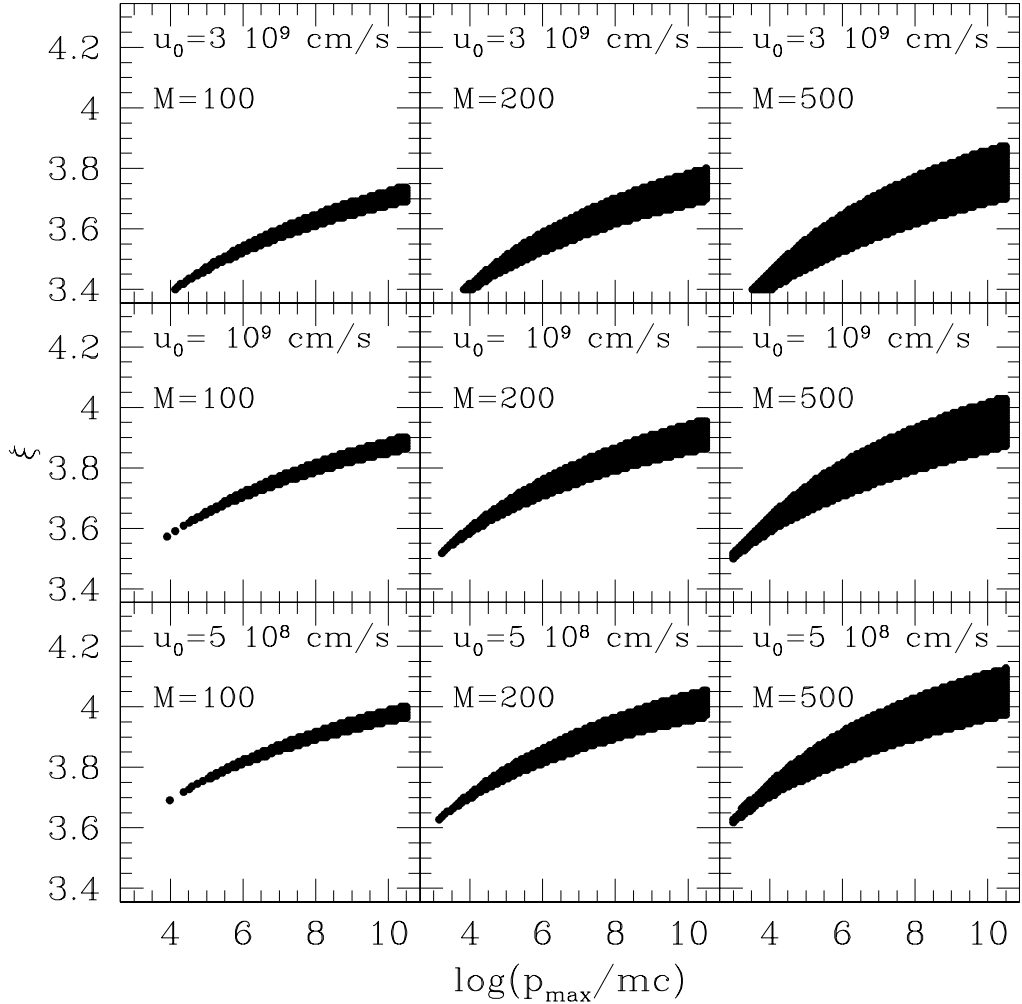}
\caption{Parameter space for multiple solutions. In the dark regions multiple solutions are still present.}
\end{figure}

\section{A recipe for injection}

The injection of particles into the cosmic ray population at a shock can be understood only considering the complex non-linear interactions between suprathermal
particles, MHD waves and background thermal plasma. Due to its intrinsic complexity, the injection process is often parametrized by
means of an injection momentum $p_{inj}$, representing the minimum momentum of the
particles that can be accelerated, and an efficiency $
\eta$, which fixes the fraction of the thermal particles that are injected in the
accelerator.
Another possibility is to adopt the {\it thermal leakage} model to describe the
injection \citep{ellisoneichler,gieseler}.
In this model, the post shock gas is assumed to be thermalized at a 
temperature $T_2$. Protons in the tail of the Maxwellian distribution can recross
 the shock and go back upstream if their velocity is high enough to allow 
 them to avoid trapping by waves. Those protons are injected in the
 accelerator. Usually, in this model, the injection momentum is set to a few times
the thermal momentum:
$
p_{inj} = \xi \sqrt{2mkT_2}
$
with $\xi$ tuned in order to fit numerical 
  \citep{gieseler} studies of diffusive shock acceleration.
  It is important to stress that the parameter $\xi$ fixes both the values of
  injection momentum and efficiency, which are no longer free parameters but are 
now connected in a physically motivated way.
It is easy to implement such a recipe in the calculations \cite{noi}.

\section{Results and conclusions}

The appearance of multiple solutions can be investigated in the 
whole parameter space, in order to define the regions where the 
phenomenon appears, when it does. In Fig. 8
we highlight the regions where there are multiple solutions (dark regions)
in a plane $\xi-\log(p_{max})$, for different values of the Mach number 
of the shock. 
The multiple
solutions disappear in most of the parameter space, and when they appear 
they look as narrow strips in the parameter space, at the boundary between regions describing
unmodified and modified shocks respectively \cite{noi}. This result may suggest that 
the narrow regions indicate the transition between two stable 
solutions, although this needs further confirmation through detailed analyses
of the stability of the solutions. 


{\it Acknowledgements:} SG acknowledges support from the Humboldt foundation.



\bibliographystyle{aipproc}   





\end{document}